\documentstyle[aps]{revtex}

\begin{document}

\draft
\title{Slope of the topological  susceptibility at zero
       temperature and finite temperature
       in the Nambu--Jona-Lasinio model}
\author{K. Fukushima\thanks{E-mail: fuku@nt1.c.u-tokyo.ac.jp},
        K. Ohnishi\thanks{E-mail: konishi@nt1.c.u-tokyo.ac.jp}
    and K. Ohta\thanks{E-mail: ohta@nt1.c.u-tokyo.ac.jp}}
\address{Institute of Physics, University of Tokyo,
         3-8-1 Komaba, Meguro-ku, Tokyo 153-8902, Japan}
\maketitle
\begin{abstract}
We estimate the slope of the topological susceptibility in the three
flavour Nambu--Jona-Lasinio model with the 't~Hooft interaction. The
results are consistent with the evaluation from the QCD sum rule in
favour of the full topological susceptibility. We apply it to the
Shore-Veneziano formula to find that it shows satisfactory agreement
with the anomalous suppression of the flavour-singlet axial charge.
The behaviour at finite temperature is also discussed.
\end{abstract}

\paragraph{Introduction.}
It is known that the ${\rm U}_{\rm A}(1)$ anomaly embodies one of the
most essential implications in the non-perturbative QCD
physics\cite{abj69}. The mass of the $\eta'$ meson is significantly
enhanced (called the ``${\rm U}_{\rm A}(1)$ problem'') and the
flavour-singlet axial charge $a^0$ is highly suppressed (called the
``spin problem'') due to the anomalous breaking of the
${\rm U}_{\rm A}(1)$ symmetry. Both phenomena are deeply connected
with another quantity sensitive to the topological characteristics of
QCD, that is, the topological susceptibility given by
\begin{equation}
 \chi(k^2)=\int{\rm d}^4x\;{\rm e}^{-{\rm i}kx}\langle0|{\rm T}
  Q(x)Q(0)|0\rangle_{\rm connected},
\label{eq:top}
\end{equation}
where $Q(x)$ is the topological charge density. The Witten-Veneziano
mass formula\cite{wit79} discloses the beautiful relation between the
quantities concerning the ${\rm U}_{\rm A}(1)$ sector, i.e.
\begin{equation}
 \frac{2N_{\rm f}}{f_\pi^2}\chi_{\rm pure}(0)=m_\eta^2+m_{\eta'}^2
  -2m_K^2,
\label{eq:w_v_formula}
\end{equation}
where $N_{\rm f}=3$ is the number of the flavours and $f_\pi$ is the
pion decay constant. This formula has been well established and
actually confirmed by the lattice calculation, which gives
$\chi_{\rm pure}(0)\sim(175\:{\rm MeV})^4$ \cite{all97} as compared
with the phenomenological value $\sim (180\:{\rm MeV})^4$ inferred
from the formula (\ref{eq:w_v_formula}). Also the flavour-singlet
axial charge could be related to the slope of the topological
susceptibility, i.e.\ $\chi'(k^2)\equiv{\rm d}\chi(k^2)/{\rm d}k^2$,
via the Shore-Veneziano formula\cite{sho95} (for a comprehensive
review see Ref.~\cite{sho98}) that is given by
\begin{equation}
 \frac{a^0(Q^2)}{a^8}=\frac{\sqrt{6}}{f_\pi}
  \left.\sqrt{-\chi_{\rm full}'(0)}\right|_{Q^2}.
\label{eq:s_v_formula}
\end{equation}
The numerical result deduced from the QCD sum rule proved to show
acceptable agreement with the EMC-SMC experiments \cite{emc88}; the
QCD sum rule calculation\cite{sho95} renders
$\left.\chi_{\rm full}'(0)\right|_{Q^2=10\,{\rm GeV}^2}=-(23.2
\;{\rm MeV})^2$, that leads to $a^0(Q^2=10\;{\rm GeV}^2)=0.353$ and
gives $\Gamma_1^{\rm p}(Q^2=10\;{\rm GeV}^2)=0.143$ for the first
moment of the polarised proton structure function.

It is worth while noting the difference between $\chi_{\rm pure}$ and
$\chi_{\rm full}$ before going on our discussion. $\chi_{\rm pure}$ is
the topological susceptibility evaluated within the pure gluonic
theory, while $\chi_{\rm full}$ is that of the full QCD. The
difference is whether the contribution from the fermionic matter
fields is contained or not. Once the theory has a massless fermion,
the vacuum becomes entirely independent of the topological theta angle
because an arbitrary ${\rm U}_{\rm A}(1)$ transformation on the
massless fermionic fields yields arbitrary shift on the theta angle.
As a result $\chi_{\rm full}(0)=0$ holds exactly in the presence of
any massless fermion. In fact non-vanishing $\chi_{\rm full}(0)$ is
attributed to the finite current quark mass. The lattice calculation
in full QCD gives $\chi_{\rm full}(0)\sim(164\:{\rm MeV})^4$ in the
region where the current quark mass is around $20\:{\rm MeV}$ with two
flavours \cite{all00}. The numerical value of $\chi_{\rm full}(0)$ is
rather close to that of $\chi_{\rm pure}(0)$. We would say, however,
that this coincidence has no convincing validity {\it a priori}. To
make this point more articulate, let us take a view of the slope of
the topological susceptibility. In the case of the pure gluonic theory
$\chi_{\rm pure}'(0)\sim(8\:{\rm MeV})^2$ is obtained from the QCD sum
rule\cite{nar91}; meanwhile, the preliminary lattice
calculation\cite{boy97} gives
$\chi_{\rm full}'(0)\sim-(19\:{\rm MeV})^2$, which is consistent with
the result from the QCD sum rule as is quoted above. Not only the
magnitude but also the sign is different between $\chi_{\rm pure}'(0)$
and $\chi_{\rm full}'(0)$. In other words the sign of the slope of the
topological susceptibility would tell us which one is regarded as
$\chi_{\rm NJL}'(0)$.

\begin{figure}
\caption{Feynman diagram to be evaluated up to the leading order of
$1/N_{\rm c}$ expansion.}
\label{fig:diag}
\caption{The behaviour of the slope of the topological susceptibility
at finite temperature}
\label{fig:slope}
\caption{The behaviour of the ratio of the axial charges at finite
temperature}
\label{fig:ratio}
\end{figure}
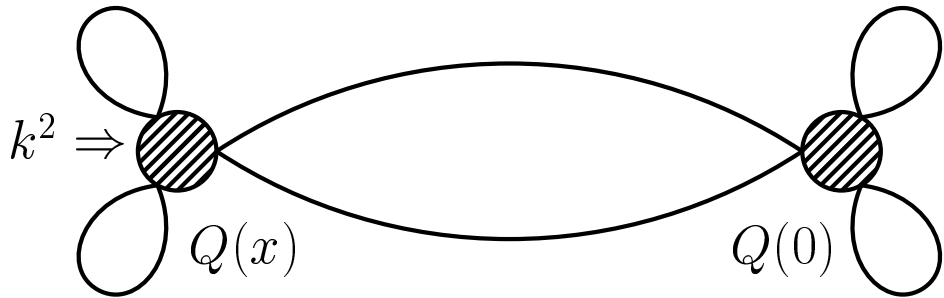
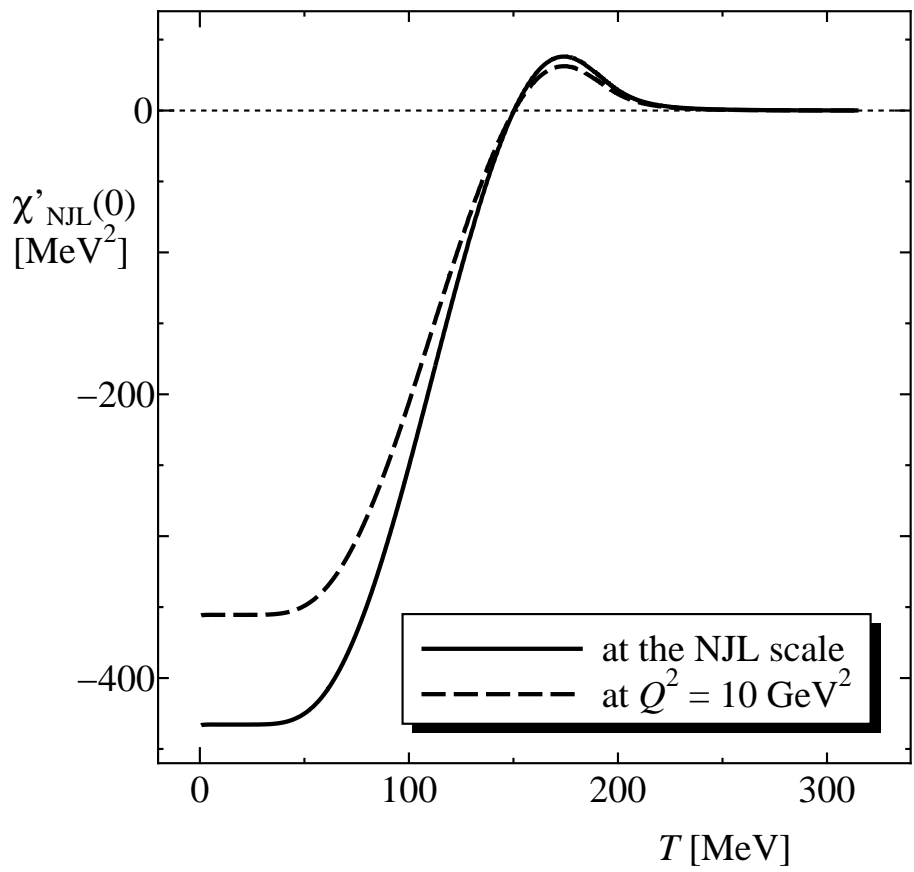
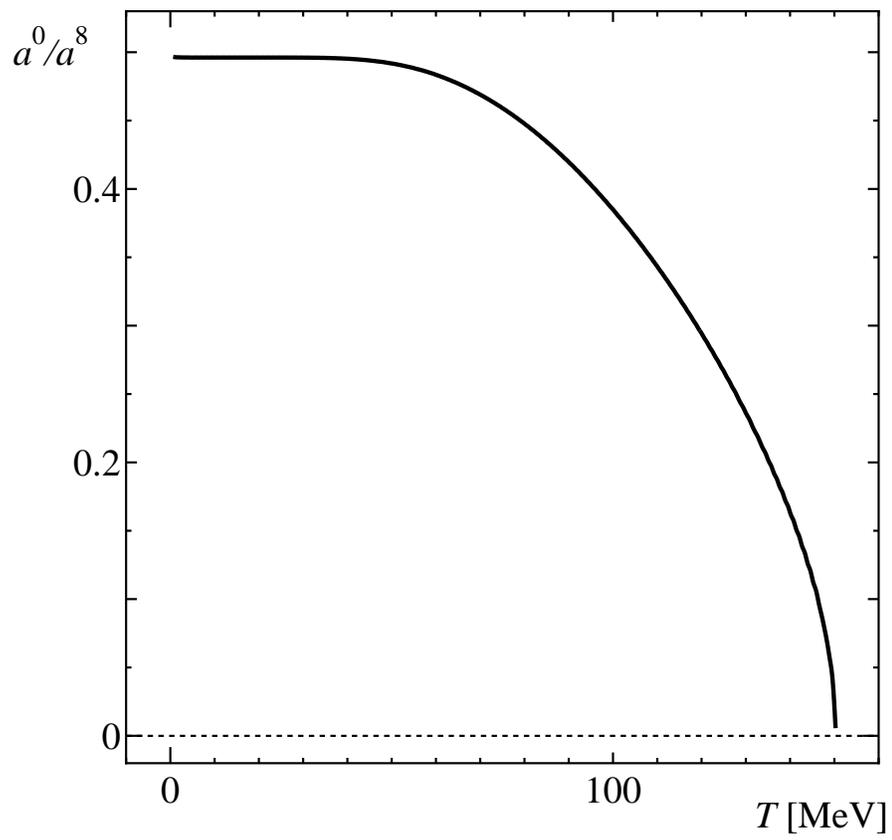

\paragraph{The slope of the topological susceptibility
 in the NJL model.}
In our previous work\cite{fuk01} we employed the topological
susceptibility to adjust the coupling strength of the 't~Hooft
interaction within the three flavour Nambu--Jona-Lasinio (NJL) model.
Although the results reproduce the desirable tendencies, there are
subtleties in the physical interpretation of the obtained topological
susceptibility in the NJL model ($\chi_{\rm NJL}(k^2)$) where the
gluonic contributions are considered to be integrated out. One of the
purposes in this letter is to clarify whether $\chi_{\rm NJL}(k^2)$
should be regarded as $\chi_{\rm full}(k^2)$ or as
$\chi_{\rm pure}(k^2)$. The topological susceptibility itself is not
sensitive to the distinction as mentioned before but the slope of the
topological susceptibility instead is suitable for this aim. Thus we
calculate $\chi_{\rm NJL}(k^2)$ with the conventional parameter set
in Ref.~\cite{hat94};
\begin{eqnarray*}
 & m_{\rm u}=m_{\rm d}=5.5\;{\rm MeV},\quad
   m_{\rm s}=135.7\;{\rm MeV},\quad
   \Lambda=631.4\;{\rm MeV}\\
 & G\Lambda^2=1.835,\quad
   K\Lambda^5=9.29
\end{eqnarray*}
for the NJL interaction Lagrangian we adopt here,
\begin{eqnarray*}
 {\cal L}_4 &=& G\sum_{a=0}^8\left[(\bar{\psi}\lambda^a\psi)^2+
  (\bar{\psi}{\rm i}\gamma_5\lambda^a\psi)^2\right],\\
 {\cal L}_6 &=& -K\left[\det\bar{\psi}(1+\gamma_5)\psi
  +\det\bar{\psi}(1-\gamma_5)\psi\right],
\end{eqnarray*}
where $\lambda^a$ is the Gell-Mann matrix in the flavour space with
$\lambda^0=\sqrt{2/3}\,{\rm diag}(1,1,1)$. The determinants are with
respect to the flavour indices. The actual procedure to calculate
$\chi_{\rm NJL}(k^2)$ is almost the same as that demonstrated in the
previous work\cite{fuk01}. The only alteration is the momentum
insertion in the expression of the topological susceptibility as shown
in Fig.~\ref{fig:diag} diagrammatically. In order to reach
$\chi'(k^2)$ we should take care in the treatment of $k^2$ because the
temporal (thermal) direction becomes distinctive at finite
temperature. In order to retain the consistency with the calculation
of $g_{\pi q\bar{q}}$ at finite temperature, the differentiation with
respect to a squared four-momentum $k^2$ is replaced by that with
respect to the fourth component $k_4^2$, which makes no difference at
zero temperature due to the Lorentz covariance.

Then the result achieved in the NJL model at zero temperature
\footnote{For further details on the calculation see
Ref.~\cite{fuk01} or our forthcoming full paper.} is
\begin{equation}
 \chi_{\rm NJL}'(0)=-(20.8\;{\rm MeV})^2,
\end{equation}
which implies that $\chi_{\rm NJL}(0)$ should be regarded as
$\chi_{\rm full}(0)$ rather than as $\chi_{\rm pure}(0)$. In fact
$\chi_{\rm NJL}(0)$ calculated with the above parameters takes the
value of $(166\;{\rm MeV})^4$\cite{fuk01}, which is close to the full
topological susceptibility achieved in the lattice simulation as is
quoted above. It follows that we should have taken
$\chi_{\rm full}(0)$ to adjust the strength of the 't~Hooft
interaction, $K$. Nevertheless, since $\chi_{\rm full}(0)$ with
massive quarks is known to behave similarly as
$\chi_{\rm pure}(0)$\cite{all00}, the essential conclusion that $K$
does not have to get smaller in order to reproduce the temperature
dependence of $\chi_{\rm pure}(0)$ would not be amended at all. So we
fix the value of $K$ at zero temperature. The resultant behaviour of
$\chi_{\rm NJL}'(0)$ at finite temperature is depicted by the solid
curve in Fig.~\ref{fig:slope}.

In order to observe the relevance upon the anomalous suppression of
the flavour-singlet axial charge it is necessary to handle
$\chi_{\rm NJL}'(0)$ at the experimental scale $Q^2=10\;{\rm GeV}^2$.
We can immediately change the scale by using the solution of the
renormalisation group equation at the one-loop order\cite{sho95};
\begin{equation}
 \left.\chi'(0)\right|_{Q^2}=\left.\chi'(0)\right|_{Q_0^2}
 \exp\left(\frac{16}{\beta_1^2\ln(Q^2/\Lambda^2)}-\frac{16}
 {\beta_1^2\ln(Q_0^2/\Lambda^2)}\right),
\end{equation}
where $\beta_1\equiv-\frac{1}{2}(11-\frac{2}{3}N_{\rm f})$ with
$N_{\rm f}=3$ and $\Lambda=350\;{\rm MeV}$ is the QCD scale
parameter. The solution is so sensitive to the initial scale parameter
$Q_0^2$ (the NJL scale) that the quantitative results would yield a
rough estimate. The dotted curve in Fig.~\ref{fig:slope} shows the
scaled solution for $Q_0^2=1\;{\rm GeV}^2$. At zero temperature the
scaled $\chi_{\rm NJL}'(0)$ is
\begin{equation}
 \left.\chi_{\rm NJL}'(0)\right|_{Q^2=10\,{\rm GeV}}=
  -(18.9\;{\rm MeV})^2,
\end{equation}
which leads to $a^0(Q^2=10\;{\rm GeV}^2)=0.288$ and
$\Gamma_1^{\rm p}(Q^2=10\;{\rm GeV}^2)=0.130$. Comparing with the
experimental results from EMC
($\Gamma_1^{\rm p}(Q^2=10.7\;{\rm GeV}^2)=0.126$), we reckon that the
description by the NJL model furnished with the 't~Hooft interaction
is quite acceptable, at least in the regime relevant to the
${\rm U}_{\rm A}(1)$ sector.

\paragraph{Summary and conclusions.}
We have evaluated the slope of the topological susceptibility within
the framework of the three flavour NJL model. Its sign implies that
$\chi_{\rm NJL}(k^2)$ should be regarded as $\chi_{\rm full}(k^2)$
rather than as $\chi_{\rm pure}(k^2)$. Also we have found that the
actual value of $\chi_{\rm NJL}'(0)$ is quite consistent with the
Shore-Veneziano formula (\ref{eq:s_v_formula}). It is the indirect
evidence not only for the validity of the interpretation of
$\chi_{\rm NJL}(k^2)$ as $\chi_{\rm full}(k^2)$ but also for the
reliability of the Shore-Veneziano formula, that is not enough tested
yet. So the results we have presented in this letter is an additional
new support for the credibility of the formula (\ref{eq:s_v_formula}).
Then we have extended our results to the finite temperature
case. $\chi_{\rm NJL}'(0)$ becomes smaller as the temperature rises
and eventually turns negative beyond $T\simeq150\;{\rm MeV}$. Thus the
Shore-Veneziano formula must break down somewhere at finite
temperature. The speed of dropping $\chi_{\rm NJL}'(0)$ is much faster
than that of the pion decay constant $f_\pi$ in the right-hand-side of
Eq.~(\ref{eq:s_v_formula}). Accordingly the ratio of the axial charges
would decrease at higher temperature as shown in Fig.~\ref{fig:ratio}.
This behaviour contradicts our naive expectation; if the effective
restoration of the ${\rm U}_{\rm A}(1)$ symmetry occurs at
sufficiently high temperature, the OZI approximation begins to work
well to give $a^0/a^8\rightarrow1$. As far as we know, the finite
temperature behaviour so far is investigated neither in experiment nor
in the lattice simulation. We suppose that our prediction shown in
Fig.~\ref{fig:slope} should be verified in other ways, say, by the
lattice calculation. We are making progress in further discussions on
the finite temperature behaviour from the phenomenological point of
view.

\acknowledgements
One of the authors (K. F. ) thanks Carl Shakin for his comments on the
full topological susceptibility.

\end{document}